\documentclass[useAMS,usenatbib,usegraphicx]{mn2e}
\title[Discovery of a strong magnetic field in the rapidly rotating B2Vn star HR~7355]{Discovery of a strong magnetic field in the rapidly rotating B2Vn star HR~7355\thanks{Based on observations obtained at the Canada-France-Hawaii Telescope (CFHT) which is operated by the National Research Council of Canada, the Institut National des Sciences de l'Univers of the Centre National de la Recherche Scientifique of France, and the University of Hawaii. } }
\author[M.E. Oksala et al.]{M.E. Oksala$^{1,2}$ \thanks{E-mail:meo@udel.edu}, G.A. Wade$^2$, W.L.F. Marcolino$^{3,4}$, J. Grunhut$^{2,5}$, 
\newauthor
D. Bohlender$^{6}$, N. Manset$^{7}$, R.H.D. Townsend$^{8}$, and the MiMeS Collaboration\\
$^{1}$Department of Physics and Astronomy, University of Delaware, Newark, DE 19716, USA\\
$^2$Department of Physics, Royal Military College of Canada, P.O. Box 17000, Station Forces, Kingston, Ontario, Canada\\
$^{3}$LAM-UMR 6110, CNRS \& Univ. de Provence, 38 rue Fr\'{e}d\'{e}ric Joliot-Curie, F-13388 Marseille cedex 13, France\\
$^{4}$Observat\`{o}rio Nacional, Rua Jos\'{e} Cristino, 77. CEP 20921-400, S\~{a}oCrist\'{o}v\~{a}o, Rio de Janeiro, Brazil  \\
$^{5}$Department of Physics, Engineering Physics \& Astronomy, Queen's University, Kingston, Ontario, Canada\\
$^{6}$National Research Council of Canada, Herzberg Institute of Astrophysics, 5071 W. Saanich Rd., Victoria, BC V9E 2E7, Canada\\
$^{7}$Canada-France-Hawaii Telescope Corporation, 65-1238 Mamalahoa Hwy, Kamuela HI 96743, USA \\
$^{8}$Department of Astronomy, University of Wisconsin-Madison, 5534 Sterling Hall, 475 N Charter Street, Madison, WI 53706, USA \\
}

\begin{document}

\date{\today}

\pagerange{\pageref{firstpage}--\pageref{lastpage}} \pubyear{2010}

\maketitle

\label{firstpage}

\begin{abstract}

We report the detection of a strong, organized magnetic field in the helium-variable early B-type star HR~7355 using spectropolarimetric data obtained with ESPaDOnS on the 3.6-m Canada-France-Hawaii Telescope within the context of the Magnetism in Massive Stars (MiMeS) Large Program.  HR~7355 is both the most rapidly rotating known main-sequence magnetic star and the most rapidly rotating helium-strong star, with $v \sin i$ = 300 $\pm$ 15 km~s$^{-1}$ and a rotational period of 0.5214404 $\pm$ 0.0000006 days.  We have modeled our eight longitudinal magnetic field measurements assuming an oblique dipole magnetic field.  Constraining the inclination of the rotation axis to be between $38^{\circ}$ and $86^{\circ}$, we find the magnetic obliquity angle to be between $30^{\circ}$ and $85^{\circ}$,  and the polar strength of the magnetic field at the stellar surface to be between 13-17 kG.  The photometric light curve constructed from HIPPARCOS archival data and new CTIO measurements shows two minima separated by 0.5 in rotational phase and occurring 0.25 cycles before/after the magnetic extrema.  This photometric behavior coupled with previously-reported variable emission of the H$\alpha$ line (which we confirm) strongly supports the proposal that HR~7355 harbors a structured magnetosphere similar to that in the prototypical helium-strong star, $\sigma$ Ori E.   
\end{abstract}


\section{Introduction}

Hot, massive stars are not expected, a priori, to exhibit magnetic fields as they lack the convective envelopes necessary for driving a dynamo.  However, Ap/Bp stars (chemically peculiar A and B type stars) are well known to possess organized magnetic fields with surface strengths up to tens of kG (e.g. Borra \& Landstreet 1979).  The hottest of the Bp stars are the so-called helium-strong stars - early B-type stars on the main sequence with enhanced and typically variable helium lines.  Some helium-strong stars show additional emission and variability in Balmer lines, variable photometric brightness and color, variable UV resonance lines, and non-thermal radio emission (Pedersen \& Thomsen 1977; Walborn 1982; Shore \& Brown 1990; Leone \& Umana 1993).  All of these quantities vary with a single period, interpreted to be the rotational period of the star.  The Balmer line (specifically H$\alpha$) and UV line variability point to the presence of a stellar wind coupled with a magnetic field to form a circumstellar magnetosphere (Shore \& Brown 1990).    Following their discovery of a magnetic field in the B2Vp star $\sigma$ Ori E (Landstreet \& Borra 1978), Borra \& Landstreet (1979) investigated eight helium-strong stars, finding six new magnetic stars.  In that paper, the authors established that a defining characteristic of the helium-strong stars is the presence of an strong, organized magnetic field. 

HR~7355 (HD~182180), the subject of the present paper, is a bright (V=6.02) helium-strong early B-type star with high $v \sin i$ (320 km~s$^{-1}$, Abt et al 2002; 270 km~s$^{-1}$, Glebocki \& Stawikowski 2000).  Originally classified as B5IV by Morris (1961), Hiltner et al. (1969) re-classified HR~7355 B2Vn, a classification that has been used subsequently.  HR~7355 has also been classified as a classical Be star (Abt \& Cardona 1984) due to observed emission in the H$\alpha$ line of its spectrum.  HIPPARCOS photometry exhibits variability with a period of $\sim$ 0.26 days (Koen \& Eyer 2002).  Rivinius et al. (2008) reported emission and variability of the H$\alpha$ profiles, as well as variability of a variety of spectral absorption lines (i.e. He, C, Si).  Those authors revisited the HIPPARCOS photometry noting that a period of 0.52 days (twice the period reported by Koen \& Eyer (2002)) could also explain the photometric variation.  Ultimately, based on the helium variability, the variable H$\alpha$ emission, and the character of the photometric variation, Rivinius et al. (2008) proposed that HR~7355 hosts a structured magnetosphere qualitatively similar to that of $\sigma$ Ori E.  With its very rapid rotation, the establishment of such a magnetosphere around HR~7355 would be of great interest.  We have therefore undertaken observations to search for the presence of a magnetic field in this star, and to further explore its spectroscopic and other physical properties.

\section{Observations}

We obtained 8 high resolution (R=65000) broadband (370-1040~nm) spectra from the spectropolarimeter ESPaDOnS attached to the 3.6-m Canada-France-Hawaii Telescope, as part of the Magnetism in Massive Stars (MiMeS) Large Program (Wade et al. 2009).  The data were obtained over the course of three nights in September 2009.  The log of observations is reported in Table \ref{magfield}.  Reduction was performed at the observatory with the Upena pipeline feeding the Libre-ESpRIT reduction package (Donati et al. 1997), which yields the Stokes I (intensity) and Stokes V (circular polarization) spectrum, as well as the diagnostic null spectrum (N), which diagnoses any spurious contribution to the polarization measurement.  Conventionally, four consecutive sub-exposures are combined to produce one polarization spectrum.  However, as each sub-exposure is 1200 seconds in duration (= 0.014 days), the combination of four sub-exposures into a single spectrum would correspond to a significant fraction (5 or 10\%) of the rotation period of this star.  We therefore processed the observations, combining only sets of two sub-exposures for a total exposure time of 2400 s.  The consequence of this change will decrease the signal to noise ratio (by a factor of $\sim \sqrt{2}$) of the spectrum, and to lose our ability to diagnose first-order systematic errors in the time domain.  The primary benefit, however, is to reduce the exposure time by half, and to thereby better resolve the rotational variation of the spectrum and magnetic field.  

Beginning on July 24, 2008, we conducted three nights (204 data points) of Johnson-Cousins V filter photometry 
of HR~7355, using the 0.9 m CTIO telescope operated by the SMARTS\footnote{The Small and Moderate Aperture Research Telescope System (http://www.astro.yale.edu/smarts/)} consortium with a $2048 \times 2046$ CCD detector.  
Variable weather plagued the first and fourth nights of observation; the second night provided no data due to weather.
A neutral density filter was employed to prevent CCD saturation. 
The observations were reduced using standard procedures (flat fielding, zero
subtraction, etc.) in  I{\sevensize RAF}.  Aperture photometry was employed to extract differential
photometric indices for HR~7355, which were then calibrated against a nearby 
comparison star, CD-28~15755.  We do not employ a photometric standard star, since we are primarily interested in changes to the
brightness of HR~7355. Cosmic rays were removed using the I{\sevensize RAF} program L{\sevensize .A.} C{\sevensize OSMIC}, a Laplacian edge detection approach described by van Dokkum (2001).
 
Additionally, archival IUE data were used to help obtain the fundamental parameters of HR~7355 (see \S3). The two IUE spectra were taken over a period of two weeks, one on Aug. 29, 1990 (H SWP 39549 L) and the other on Sept. 7, 1990 (H SWP 39596 L).  Both observations were taken using the IUE large aperture.

\begin{table}
\caption{Summary of ESPaDOnS observations and magnetic field measurements.  Column 1 gives the heliocentric Julian date of mid-observation.  Column 2 the rotational phase according the adopted ephemeris given in Eq. (3).  Column 3 the peak signal to noise ratio per 1.8 km~s$^{-1}$ per spectral pixel in the reduced spectrum.  Column 4 the signal to noise ratio in the LSD profile. Columns 5 and 6 the longitudinal magnetic field and its associated error.}
\begin{tabular}{lccccc}
\hline \hline
HJD & Phase & Peak S/N & LSD S/N  & $B_\ell$ & $\sigma_B$ \\ 
(2450000+)  &   & & & (G) & (G) \\ \hline
5077.81898 & 0.75 & 902 & 21162 &  $-$1992 & 120 \\
5077.84782 & 0.81 & 928 & 22000 &  $-$1879 & 123 \\
5081.75435 & 0.30 & 870 & 20041 &  +2501 & 130 \\
5081.78316 & 0.35 & 724 & 18198 &  +2471 & 162 \\
5081.81512 & 0.41 & 792 & 18927 &  +2116 & 164 \\
5081.84393 & 0.47 & 568 & 15278 &  +1306 & 234 \\
5082.75151 & 0.21 & 1072 & 23605 &  +2372 & 106 \\
5082.78033 & 0.26 & 1122 & 23753 &  +2344 & 101 \\
\hline \hline
\end{tabular}
\label{magfield}
\end{table}

\section{Stellar Parameters}

In order to determine the photospheric parameters of HR~7355 we used  
a comprehensive grid of metal line blanketed, non-LTE models from  
TLUSTY (Lanz \& Hubeny 2007) and also the SYNSPEC line formation code 
\footnote{http://nova.astro.umd.edu/Synspec43/synspec.html}. Effective  
temperatures ($T_{\rm{eff}}$) from 15000~K to 22000~K (steps of 1000  
K) and surface gravities (log $g$) from 3.0 to 4.0 (steps of 0.25 dex)  
were considered to model the observed optical and UV spectra\footnote{
We chose to focus our analysis on the optical spectrum corresponding
to phase 0.47, where we have the lowest degree of contamination from circumstellar
material (see Section 4). We note however that the physical parameters derived 
represent well the other spectra within the error bars.}.

The spectral type and the high rotational velocity presented by  
HR~7355 ($v \sin i $ = 300 $\pm$ 15 km~s$^{-1}$) hindered the use of  
silicon lines (e.g., Si~{\sc ii} 412.8-413.1~nm and Si~{\sc iii} 
455.3-457.5~nm) to determine $T_{\rm{eff}}$. Alternatively, the  
intensity of the Balmer and He~{\sc i} transitions were found to be  
good diagnostics (see for example, Dufton et al. 2006).  For the  
determination of $\log g$, we used mainly H$\gamma$. Our best  
fit for $T_{\rm{eff}}$ was 17000~K, and for $\log g$ the best value  
was 3.75.   After the comparison of several grid models to the  
observed spectrum, we estimated the errors in $T_{\rm{eff}}$ and log  
$g$ to be about $\pm $2000~K and $\pm 0.25$~dex, respectively. 
These  conservative errors were adopted mainly due to the
uncertainties involved in the normalization of our echelle spectra, the observed
spectral variability, and the fast rotation presented by this star, which causes 
a large broadening and blends several features, complicating the
analysis.

The physical parameters derived are summarized in Table \ref{params}.
Very good agreement is obtained in both UV and optical regions. Since
TLUSTY directly provides only $T_{\rm{eff}}$ and log $g$, we have
computed the luminosity, radius, and the mass of HR~7355 in the
following way. First, we computed the visual extinction $A_V$ from the
apparent (V = 6.02) and ISM-corrected visual magnitudes (V$_0 = 5.81$;
Guti\'errez-Moreno \& Moreno 1968). Thereafter, we inferred the
distance from the HIPPARCOS parallax for this object ($\pi = 3.66 \pm
0.37$; van Leeuwen 2007) and computed the absolute visual magnitude
($M_V$). By using the bolometric correction (BC) corresponding to our TLUSTY
model (Lanz \& Hubeny 2007), we thus determined the bolometric
magnitude ($M_{\rm{BOL}} = M_V + BC$) and then the luminosity, taking
the bolometric magnitude of the Sun to be 4.75 (Allen 1976).  Taking
the distance and its associated uncertainty, we find that $\log L_\star/L_
\odot = 3.06^{+0.09}_{-0.08}$. From this value and $T_{\rm{eff}}$, we
computed the radius and from log $g_{\rm{true}}$ (see below) we derived
the mass. These parameters are in good agreement with typical early B  
star values (see e.g. Trundle et al. 2007).  Note that if we use the
$uvby\beta$ colors of HR~7355 to derive $T_{\rm{eff}}$, we obtain a
value of $\sim$ 18000~K, in agreement with our model analysis.
Regarding log $g$, the inferred value of 3.75 should be considered as
effective, i.e., the surface gravity reduced by rotation. An estimate
of the true gravity can be obtained by the following expression
(Repolust et al. 2004):
\begin{equation}
\log g_{\rm{true}} =  \log (g + (v \sin i)^2/ \rm{R}_{\star}).
\end{equation}
\noindent For HR~7355, we find $\log g_{\rm{true}} = 3.95$, which
is more consistent with its spectral classification. The stellar mass
follows from this value (see Table \ref{params}).

As previously claimed in the literature, we found that this star must
be enriched in helium. We could not fit any of the observed He~{\sc i}  
profiles using the solar abundance number ratio He/H = 0.1. All the  
computed lines were systematically weaker than the ones observed.  
Models with He/H ratios from 0.2 to 1.0 presented good agreement with  
the observations (taking into account the different spectra). 
A compatible He/H ratio (0.4) was derived by Rivinius et al. (2008)  
from equivalent widths of He I lines in a spectrum acquired in 1999.   
However, these same authors estimated a solar He/H value from a second
spectrum, obtained in 2004.

\begin{table}
\caption{Summary of photospheric properties of HR~7355. The
determination of the
rotational period $P_{\rm{rot}}$ is discussed in Section 6.}
\begin{tabular}{ll}
\hline
Spectral type             & B2Vn              \\
$T_{\rm{eff}}$ (K)        & 17 000 $\pm$ 2000 \\
log $g_{\rm{true}}$ (cgs) & 3.95 $\pm$ 0.25    \\
$\xi_t$ (km~s$^{-1}$)     &  2                \\
He/H                     & 0.2-1.0 \\
\\
log L$_\star$/L$_\odot$    & 3.06 $\pm ^{0.09}_{0.08}$ \\
R$_{\star}$/R$_\odot$      & 3.9 $\pm ^{1.1} _{0.8}$ \\
M$_{\star}$/M$_{\odot}$    & 5.0 $\pm ^{6} _{3}$ \\
\\
$v \sin i$ (km~s$^{-1}$)     & 300 $\pm$ 15                \\
$P_{\rm{rot}} (d)$ & 0.5214404 $\pm$ 0.0000006  \\
\hline
\end{tabular}
\label{params}
\end{table}

\section{Magnetic Field}

\begin{figure*}
\centering
\includegraphics[width=170mm]{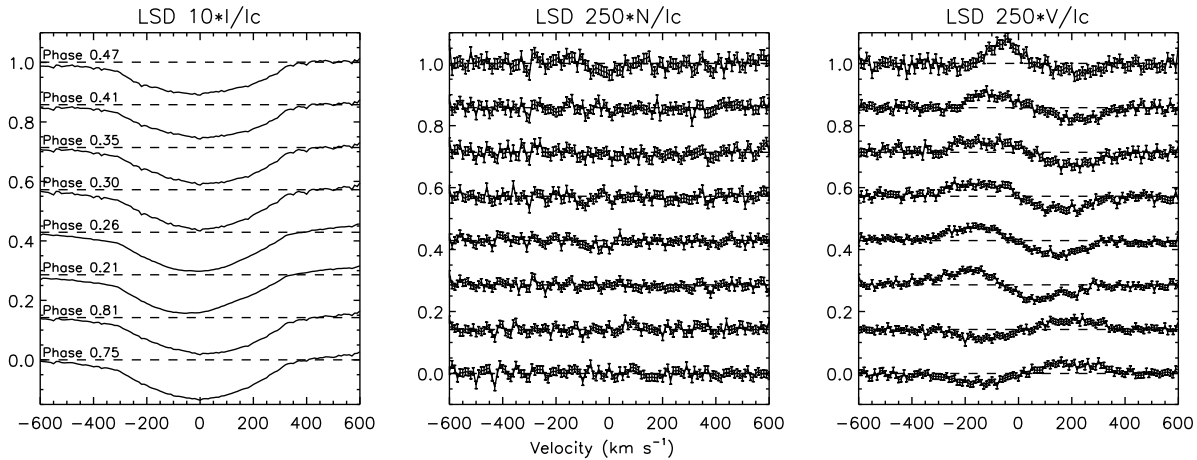}
\caption{LSD Stokes~$I$ (left), null $N$ (middle), and Stokes~$V$ (right) profiles of HR~7355. The profiles are expanded by the indicated factor and each separate observation is shifted upwards for display purposes. The phase noted for each Stokes $I$ profile is calculated using the ephemeris in \S6 and and are the same for the corresponding $N$ and Stokes $V$ profiles.  Clear Zeeman signatures are detected in each Stokes~$V$ profile, while the null profile shows no signal.  For visualization purposes, these profiles were binned to 10 km~s$^{-1}$ in wavelength.}
\label{LSDprof}
\end{figure*}

We used the multiline analysis method of Least-Squares Deconvolution (LSD; Donati et al. 1997) to produce mean Stokes I and V profiles from our spectra. These profiles are illustrated in Fig. \ref{LSDprof}. The details of the LSD method as applied to B stars are described by Silvester et al. (2009).  The LSD line mask was created using a VALD {\sc extract stellar} line list for a star with $T_{\rm{eff}}$=18000~K, log $g$ = 4.0, and solar abundances except for helium.  The helium abundance was set to $-$0.1 in logarithmic units, although as mentioned previously, the helium line strength varies strongly with rotation phase.  Most helium and metal lines in the range 405-800~nm were included in the line mask.  Lines located in regions with Balmer lines were removed as hydrogen lines are not included in the mask.  The wavelength range was chosen in an attempt to exclude regions with strong blending with hydrogen lines in the blue-most part of the spectrum and strong telluric contamination in the red part of the spectrum.  Mean profiles were binned to 2.6 km~s$^{-1}$ in wavelength.    The LSD velocity range was set from $-$600 km~s$^{-1}$ to +600 km~s$^{-1}$ and the line depth cutoff was 10\% of the continuum.  The mean Stokes I, null, and mean Stokes V profiles are shown in Fig. \ref{LSDprof}.  The lack of signal in the null profiles indicate that there are no important spurious contributions to the Stokes V profiles.

Each separate LSD Stokes V profile produced a definite detection ($>$ 99.999\%) according to the criteria described by Donati et al. (1997).  From the LSD mean Stokes I and V profiles, we have calculated the longitudinal magnetic field from the first moment of Stokes V as a function of velocity $v$ from: 
\begin{equation}
B_{\ell} = -2.14 \times 10^{11} \frac{ \int v V(v) dv}{\lambda g c \int [1-I(v)]dv}
\end{equation}
\noindent (Mathys 1989; Donati et al. 1997; Wade et al. 2000).  In Eq. (2), $g$ is the mean Land\'{e} factor and $\lambda$ is the mean wavelength of all the included lines.  The integration range employed in calculating the longitudinal magnetic field was from $-$350 km~s$^{-1}$ to +350 km~s$^{-1}$.   The derived values, along with their uncertainties, are reported in Table \ref{magfield}.  Typical uncertainties are 100-150 G.  The longitudinal magnetic field is observed to change from about $-$2 kG to +2.5 kG.  Such a strong longitudinal field suggests an organized magnetic field with a likely surface dipole component stronger than 8~kG.

\section{Rotation Period}

\begin{figure}
\centering
\includegraphics[width=3.5in]{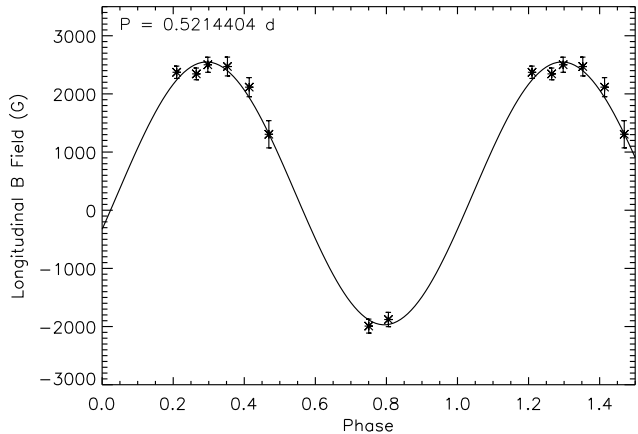}
\caption{Longitudinal magnetic field measurements for HR~7355 as reported in Table \ref{magfield}, and the best-fit first-order sine curve. The data are phased using a period of 0.5214404 days and the same JD$_{0}$ in \S6.  Plotting the data with a period of 0.2607202 days does not phase the data well.}
\label{fullper}
\end{figure}

The rotation period of HR~7355 remains unclear following the analysis of original HIPPARCOS data by Koen \& Eyer (2002).  The proposed ``single minimum'' period of approximately 0.26 d would seem, from the rotational velocity, to imply that the star is rotating more rapidly than the critical (break-up) rotation rate.  Unfortunately, our new photometry from the 0.9m telescope at CTIO does not resolve this ambiguity, as both the shorter period and the longer period ($\sim$ 0.52~d) produce reasonable light curves.  However, when the longitudinal magnetic field measurements are plotted with both periods, the longitudinal magnetic field cannot be phased with the 0.26 day period in a reasonable fashion.  However, when phased with the 0.52 day period the longitudinal field measurements describe a smooth, approximately sinusoidal variation from $-$2~kG to +2.5~kG (Fig. \ref{fullper}).

\begin{figure}
\centering
\includegraphics[width=3.5in]{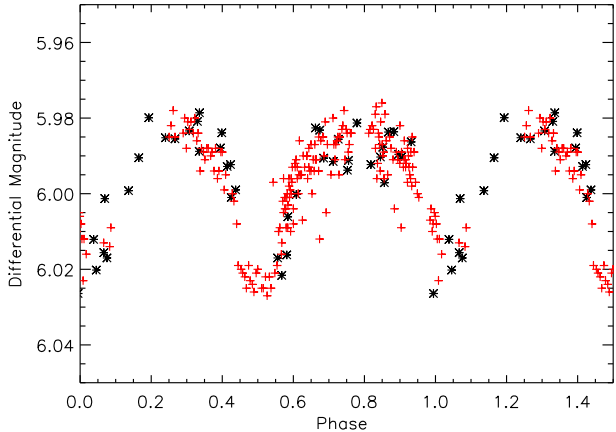}
\caption{The V-band photometric light curve for HR~7355 phased using the ephemeris of \S6.  The HIPPARCOS photometry are plotted as black asterisks, while the new CTIO data (see \S2) are plotted as red plus signs.}
\label{photLC}
\end{figure}

A periodogram of the combined HIPPARCOS and CTIO photometry, in the region near the HIPPARCOS solution, gives two closely spaced periods, 0.5214184 d and 0.5214404 d.  The longer period is consistent with the periods derived by both Mikul\'{a}\u{s}ek et al. (2010) and Rivinius et al. (2010).  Therefore, we tentatively select the longer period of \textbf{0.5214404 $\pm$ 0.0000006 d} and a photometric light curve consisting of \textbf{two separate minima}.  We define a new ephemeris for HR~7355:
\begin{equation}
JD = (2454672.79 \pm 0.01) + (0.5214404 \pm 0.0000006) \cdot E.
\end{equation}
\noindent The JD$_{0}$ is determined from one of the minima in the photometric light curve obtained from the new CTIO data.  Both sets of photometry are plotted and phased in Fig. \ref{photLC}, according to this new ephemeris.   We computed the periodogram of the eight magnetic field measurements and obtained a best fit period of 0.523 $\pm$ 0.006 d, consistent with the period found by photometry.  When the $B_\ell$ measurement are phased according to Eq. (3), the magnetic extrema occur at phases $\sim$0.3 and $\sim$0.8.

\section{Discussion}

The sinusoidal variation of the longitudinal magnetic field indicates an important dipole component to the stellar magnetic field.  Using the characteristics of the sinusoidal fit to the magnetic data and the stellar parameters with their error bars (derived in \S3), we can estimate the geometry and strength of the dipole: the inclination of the rotational axis with respect to the line of sight to the observer, $i$, the obliquity of the dipole axis relative to the rotational axis, $\beta$, and the polar strength of the surface dipole field, $B_{d}$, as described by e.g. Wade et al. (1997).  We have determined the allowed range of inclinations $38^{\circ} \le i \le 86^{\circ}$, assuming rigid rotation.   Lower inclinations were not feasible as the implied equatorial rotation velocity (i.e. $v \sin i / \sin i$) was higher than the critical break-up velocity (v$_{\rm{crit}}$; Townsend et al. 2004) calculated for the mass and radius considered.  Using this range of $i$ and the observed values of the magnetic extrema (Fig. \ref{fullper}, lower panel),  we calculate the  obliquity to be $30^{\circ} \le \beta \le 85^{\circ}$.  With this inclination and obliquity, and the maximum longitudinal field, we compute that the surface dipolar field strength for HR~7355 is 13-17 kG at the pole.  

The Rigidly Rotating Magnetosphere (RRM) model was proposed by Townsend \& Owocki (2005) to describe a rapidly rotating star in which the magnetic field overpowers the stellar wind, allowing plasma to become trapped in a magnetosphere that co-rotates along with the star.  The geometry of the magnetosphere depends on the obliquity angle and the rotational velocity of the star (See Townsend 2008).  The variability of the observables depends on the inclination angle.  This model was applied to the prototypical helium-strong star, $\sigma$ Ori E, by Townsend et al. (2005).  Comparing our picture of the observations and geometry of HR~7355 with the observations and model of $\sigma$ Ori E, the two show similarities in both the variability of observables, but also in the relative timing of each observation related to the geometry.  In both stars, the minimum of the photometric light curve corresponds to minimum emission in H$\alpha$, as well as a null in the longitudinal field curve.  Although the photometric light curves both show similar V-band magnitude changes ($\sim$0.05 mags), the durations of the photometric ``eclipses'' in HR~7355 are slightly longer than those in $\sigma$ Ori E; this is not surprising given the more rapid rotation of HR~7355.  While the phasing of the H$\alpha$ variations and the photometric minima of HR~7355 are consistent with circumstellar material, the photometric variations may also be due to photospheric spots on the surface of the star.  Based on this small initial dataset, the RRM model appears to be capable of explaining the basic observational phenomena presented by HR~7355.  We therefore expect it to be useful to confirm and characterize the magnetosphere of HR~7355 once more data are acquired.  

HR~7355 is both the most rapidly rotating helium-strong star and the most rapidly rotating magnetic star discovered thus far with a 0.5214404 $\pm$ 0.0000006 day period and a $v \sin i$ of 300 km~s$^{-1}$, surpassing the Bp star CU Vir ($P_{\rm{rot}}$ = 0.52070308 d, $v \sin i$ = 160 km~s$^{-1}$, Kuschnig et al. 1999).  HR~7355 has a strong magnetic field ($B_{d} \sim$ 13-17~kG) that when coupled with its wind produces metal line variability and a magnetosphere that co-rotates with the star, resulting in variability of H$\alpha$ and, photometric brightness.   HR~7355 is rotating near its critical velocity, making it an extreme laboratory to study the effects of a magnetosphere under these conditions, and perhaps even to provide a link between Bp and Be stars.  However, the high rotation of this star also creates challenges for spectral line analysis.  At rotational velocities near critical, the star becomes oblate and gravity darkening effects become important.  When viewed from the equator, the star can appear reddened and the temperature and surface gravity are lowered (Townsend et al. 2004).  Detailed line-profile modeling is required to determine accurate parameters.  Although historically HR~7355 is classified as B2Vn, the temperature derived from its spectrum suggests a later type.  In addition to continuing to monitor the magnetic field, line profile and photometric variations of HR~7355, we intend to undertake, in a future study, a more physically realistic line profile analysis, taking into account rotational deformation and gravity darkening, to provide a consistent solution for stellar parameters and evolution.

\section*{Acknowledgments}
GAW acknowledges support from a Natural Science and Engineering Research Council of Canada (NSERC) Discovery Grant and a Department of National Defense ARP grant.  RHDT acknowledges support from NASA / LTSA grant NNG05GC36G.

\label{lastpage}

\end{document}